\documentclass[a4paper,aps,preprintnumbers,showpacs,superscriptaddress,nofootinbib,amsmath,amssymb]{revtex4}
\usepackage{graphicx}
\usepackage{hyperref}
\usepackage{subfigure}
\usepackage{multirow}
\usepackage[toc,page]{appendix}
\usepackage[utf8]{inputenc}
\usepackage{ulem}
\DeclareMathAlphabet{\pazocal}{OMS}{zplm}{m}{n}

\begin{document}
\title{Quasinormal modes of Dirac field in the Einstein-dilaton-Gauss-Bonnet and Einstein-Weyl gravities}
\author{A. F. Zinhailo}\email{F170631@fpf.slu.cz}
\affiliation{Institute of Physics and Research Centre of Theoretical Physics and Astrophysics, Faculty of Philosophy and Science, Silesian University in Opava, CZ-746 01 Opava, Czech Republic}

\begin{abstract}
Quasinormal modes of Dirac field in the background of a non-Schwarzschild black holes in theories with higher curvature corrections are investigated in this paper. With the help of the semi-analytic WKB approximation  and further  using  of Padé approximants  as prescribed in \cite{Matyjasek:2017psv} we consider quasinormal modes of a test massless Dirac field in the Einstein-dilaton-Gauss-Bonnet (EdGB) and Einstein-Weyl (EW) theories. Even though the effective potential for one of the chiralities has a negative gap we show that the Dirac field is stable in both theories. We find the dependence of the modes on the new dimensionless parameter $p$ (related to the coupling constant in each theory) for different values of the angular parameter $\ell$ and show that the frequencies tend to linear dependence on $p$. The allowed deviations of qausinormal modes from their Schwarzschild limit are one order larger for the Einstein-Weyl theory than for the Einstein-dilaton-Gauss-Bonnet one, achieving the order of tens of percents. In addition, we test the Hod conjecture which suggests the upper bound for the imaginary part of the frequency of the longest lived quasinormal modes by the Hawking temperature multiplied by a factor. We show that in both non-Schwarzschild metrics the Dirac field obeys the above conjecture for the whole range of black-hole parameters.
\end{abstract}
\maketitle

\section{Introduction}\label{introduction}

Recently, the interest in studying new alternative theories of gravity has been increasing see, for instance, \cite{alternative}. In spite of its efficiency, the unmodified general theory of relativity is not able to answer some fundamental questions. Some of the most important problems are the construction of non-contradictory quantum gravity, the singularity problem, the problems of dark matter and dark energy. The problem of the construction of non-contradictory quantum gravity is connected with the non-renormalizability of General Relativity. This can usually be solved by adding the higher order terms in curvature to the theory \cite{tHooft}. In this paper, we will consider two different approaches: the first approach is related to adding of the Gauss Bonnet term coupled to a dilaton \cite{Blazquez-Salcedo:2016enn,Pani:2009wy,Nampalliwar:2018iru}, while the second theory consists of the Weyl term \cite{Einstein-Weyl:2018pfe} added to the Einstein action. Both theories are inspired by the low energy limit of string theory \cite{low-energy}, which contain quadratic corrections in curvature, but the Gauss-Bonnet term alone leads to the full divergence and does not contribute to the equations of motions, so that there remaining only two options for adding higher curvature corrections: either coupling of the Gauss-Bonnet term to other fields or choosing essentially non-Gauss-Bonnet quadratic corrections. Thus, here we will consider example of the both options.

The Lagrangian of the Einstein-dilaton-Gauss-Bonnet gravity is:
\begin{eqnarray} \label{lagranzianEdGB}
{\cal L}_{EdGB}&=&\frac{1}{2}R - \frac{1}{4} \partial_\mu \phi \partial^\mu \phi \\\nonumber&&+ \frac{\alpha '}{8g^2} e^{\phi }\left(R_{\mu\nu\rho\sigma}R^{\mu\nu\rho\sigma} - 4 R_{\mu\nu}R^{\mu\nu} + R^2\right),
\end{eqnarray}
where $\alpha '$ is the Regge slope, $g$ is the gauge coupling constant and $\phi$ is the dilaton field function. Black holes in the Einstein-dilaton-Gauss-Bonnet gravity has been recently investigated in  number of papers \cite{Nampalliwar:2018iru}, \cite{Ayzenberg:2014aka,Maselli:2014fca,Maselli:2015tta,Cunha:2016wzk,Konoplya:2016jvv,Zhang:2017unx,Prabhu:2018aun,Nair:2019iur,Konoplya:2019hml}.

For the Einstein-Weyl gravity the Lagrangian can be written as follows:
%
\begin{eqnarray} \label{lagranzianEW}
{\cal L}_{EW}&=&\sqrt{-g} (\gamma R - \alpha C_{\mu\nu \rho\sigma} C^{\mu\nu \rho\sigma} + \beta R^2),
\end{eqnarray}
where $\alpha$, $\beta$ and $\gamma$ are coupling constants, $C_{\mu\nu\rho\sigma}$ is the Weyl tensor. For spherically symmetric and asymptotically flat solutions we can choose $\gamma =1$ and $\beta =0$ \cite{EW}, so that the only new coupling constant is $\alpha$. The condition $R=0$ is evidently satisfied in this case, so that the Schwarzschild solution is also the solution of the above theory. The static spherically symmetric and asymptotically flat black holes in the Einstein-Weyl theory represent the generic class of black hole solutions in the quadratic theories of gravity if no other matter fields are added. They have been recently studied in \cite{Lin:2016kip,Zinhailo:2018ska,Konoplya:2019ppy,Wu:2019uvq}.

Recently black holes in the both theories have been extensively studied. In particular, quasinormal modes were found for test scalar and electromagnetic fields \cite{Zinhailo:2018ska,Konoplya:2019hml}. Although  quasinormal modes of a Dirac field around black holes in the Einstein gravity were studied in detail in a number of papers (see \cite{Cho:2003qe,Jing:2003wq,Giammatteo:2004wp,Jing:2005dt,Blazquez-Salcedo:2018bqy} and reference therein), to the best of our knowledge there are no works devoted to Dirac quasinormal modes in theories with higher curvature corrections. When considering the neutrino field, the special attention must be paid to the presence of a negative region of a potential curve with negative chirality. The positive definite effective potential guarantees dynamical stability of perturbations, that is, absence of unboundedly growing modes. For the one of the chiralities of Dirac field in the Schwarzschild background the negative gap does not lead to the instability because the other chirality provides positive definite potential and the both chiralities are proved to be iso-spectral. However, the iso-spectrality has never been proved for the considered non-Einsteinian theories, so that the instability cannot be excluded a priori. Because of this, it would be interesting to study the quasinormal spectrum of the Dirac field in the above non-Einsteinian theories of gravity and see whether there is an instability. After all, the test of stability is extremely important for higher curvature corrected theories because of the so called eikonal instability which occurs in a abroad class of theories with various higher curvature corrections, spacetime dimensions and asymptotics  \cite{Dotti:2005sq,Gleiser:2005ra,Takahashi:2010gz,Grozdanov:2016fkt,Konoplya:2017zwo,Cuyubamba:2018jdl}, and not only for gravitational, but also for test fields \cite{Gonzalez:2017gwa}.

We will analyze values of modes at the low angular parameter $\ell$  and in the eikonal regime. We will find dependencies of the complex frequency on the new dimensionless parameter $p$ (related to the coupling constant in each theory). In addition, we will compare quasinormal modes of both theories between each other and with modes of other fields in each theory. In addition here we will test the quasinormal modes of Dirac field in the above two theories as to the Hod's conjecture \cite{Hod:2006jw} who claims that there must always be a minimal mode whose damping rate is limited by the Hawking temperature multiplied by some factor.

This work is organized as follows. In Sec.~\ref{sec:metricsection} we introduce a metric and a general wave equation and consider the effective potential for the Dirac field for a spherically symmetric black hole. For this case we prove that the Dirac perturbations are linearly stable in both theories. In Sec.~\ref{sec:massless}, the basic principles of the WKB method are briefly considered, an analytical approximation in the eikonal regime is analysed Sec.~\ref{sec:eikonal}, the quasinormal modes for test massless Dirac field are found, a comparative analysis is made for the our result with the results for other fields in these theories of gravity Sec.~\ref{sec:lsmall}. In Sec.~\ref{sec:hods} we will check the Hod's conjecture for the Dirac field in Einstein-dilaton-Gauss-Bonnet and Einstein-Weyl gravities.

\section{Black hole metric and analytics for the wave equation}\label{sec:metricsection}
In the general case the metric for a spherically symmetric black hole can be written in the form:
%
\begin{eqnarray}\label{metric}
ds^2 &=& -e^{\mu(r)}dt^2+e^{\nu(r)}{dr^2}+r^2 (\sin^2 \theta d\phi^2+d\theta^2),
\end{eqnarray}
where $e^{\mu(r)}$ and $e^{\nu(r)}$ are the metric coefficients. The explicit expression for the metric coefficients were obtained numerically in \cite{Kanti:1995vq} for Einstein-dilaton-Gauss-Bonnet gravity and in \cite{EW} for Einstein-Weyl gravity. The approximate analytical expressions (which will be used here) were obtained in \cite{Kokkotas:2017ymc} for the Einstein-dilaton-Gauss-Bonnet metric, in \cite{Kokkotas:2017zwt} for the Einstein-Weyl metric. They are also written down in Appendixs \ref{Appendix1}, \ref{Appendix2}.

We parameterize the both black-hole solutions  in theories (\ref{lagranzianEdGB}, \ref{lagranzianEW}) via the following dimensionless parameter $p$ up to the rescaling:
%
\begin{subequations}\label{parameter}
\begin{eqnarray}\label{parameterpedgb}
p_{EdGB}\equiv6e^{2\phi_0}=\frac{6\alpha'^2}{g^4r_0^4}e^{2(\phi_0-\phi_{\infty})} \qquad{(Einstein-dilaton-Gauss-Bonnet)}\,
\end{eqnarray}
\begin{eqnarray}\label{parameterpew}
p_{EW}=\frac{r_0}{\sqrt{2\alpha}} \qquad{(Einstein-Weyl)}.
\end{eqnarray}
\end{subequations}

For convenience we fix radius of the black-hole event horizon to be $r_0=1$. For all $p$ the Schwarzschild metric is the exact solution of the Einstein-Weyl equations as well, but only at some minimal nonzero $p_{min}$, in addition to the Schwarzschild solution, there appears the non-Schwarzschild branch which describes the asymptotically flat black hole, whose mass is decreasing, when $p$ grows. The approximate maximal and minimal values of $p$ are:
%
\begin{subequations}\label{paraneter}
\begin{eqnarray}\label{parameteredgb}
p_{min,EdGB} \geq 0, \quad p_{max,EdGB}\leq0.97,
\end{eqnarray}
\begin{eqnarray}\label{parameterew}
p_{min,EW} \approx 1054/1203 \approx 0.876, \quad p_{max,EW} \approx 1.14.
\end{eqnarray}
\end{subequations}

The general covariant Dirac equation has the form \cite{Brill:1957fx}:
\begin{equation}\label{covdirac}
\gamma^{\alpha} \left( \frac{\partial}{\partial x^{\alpha}} - \Gamma_{\alpha} \right) \Psi=0,
\end{equation}
where $\gamma^{\alpha}$ are noncommutative gamma matrices and $\Gamma_{\alpha}$ are spin connections in the tetrad formalism. We separate of angular variables in equation (\ref{covdirac}) and rewrite the wave equation in the following general master form in terms of the ``tortoise coordinate'' $r_*$ \cite{Brill:1957fx}:
\begin{equation}  \label{klein-Gordon}
\dfrac{d^2 \Psi}{dr_*^2}+(\omega^2-V(r))\Psi=0, \quad dr_*=\sqrt{e^{\nu(r)-\mu(r)}}dr.
\end{equation}

The effective potentials of test Dirac ($s=\pm 1/2$) field in the general background (\ref{metric}) can be written as follows:
\begin{equation}
V_{\pm}(r) = \frac{k}{r}\left(\frac{e^{\mu(r)} k}{r}\mp\frac{e^{\mu(r)}\sqrt{e^{\nu(r)}}}{r}\pm\sqrt{e^{\mu(r)-\nu(r)}}(\sqrt{e^{\mu(r)}})'\right),
\end{equation}
where the prime designates the differentiation with respect to the ``tortoise coordinate'' $r_{*}$.
 \vspace*{0.5em plus .6em minus .5em}
\begin{figure}[ht]
\vspace{-4ex} \centering \subfigure[]{
\includegraphics[width=0.45\linewidth]{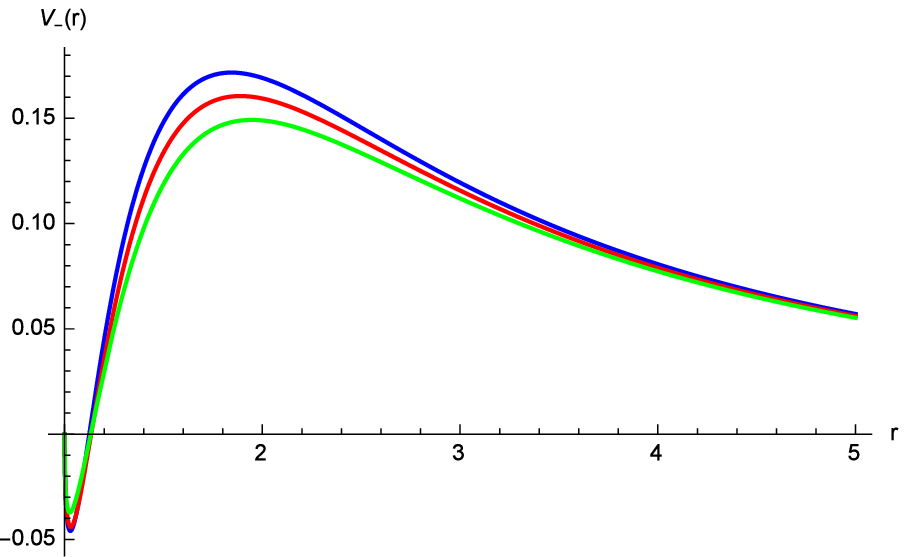} \label{fig:EdGB1_a} }
\hspace{4ex}
\subfigure[]{
\includegraphics[width=0.45\linewidth]{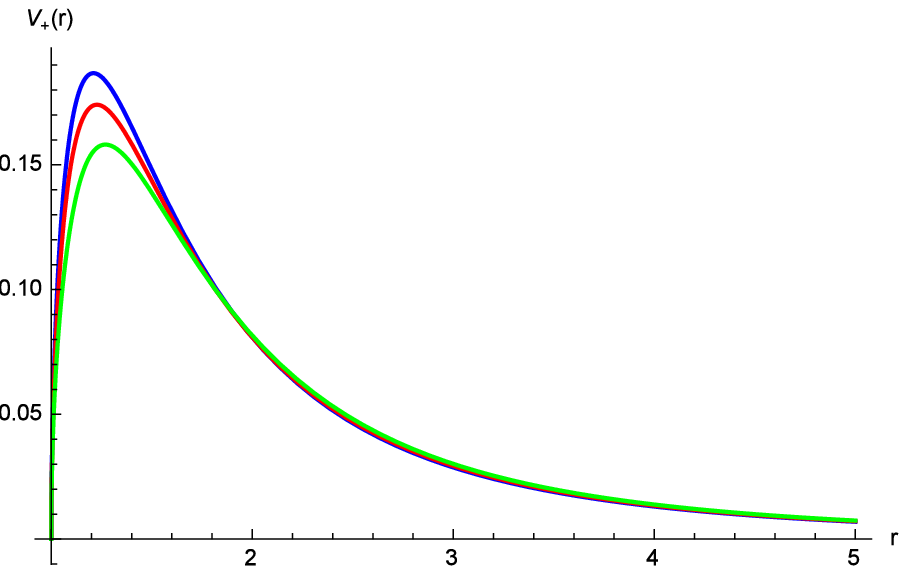} \label{fig:EdGB2_b} }
\caption{The effective potential $V(r)$ for the EdGB gravity for $\ell=1$; the blue line
corresponds to $p=0$, the red line corresponds to $p=0.5$ and the green line is $p=0.97$:
 \subref{fig:EdGB1_a} $V_{-}(r)$;
 \subref{fig:EdGB2_b} $V_{+}(r)$.} \label{fig:vedgb}
 \label{ris:one1}
\end{figure}
\begin{figure}[ht]
\vspace{-4ex} \centering \subfigure[]{
\includegraphics[width=0.45\linewidth]{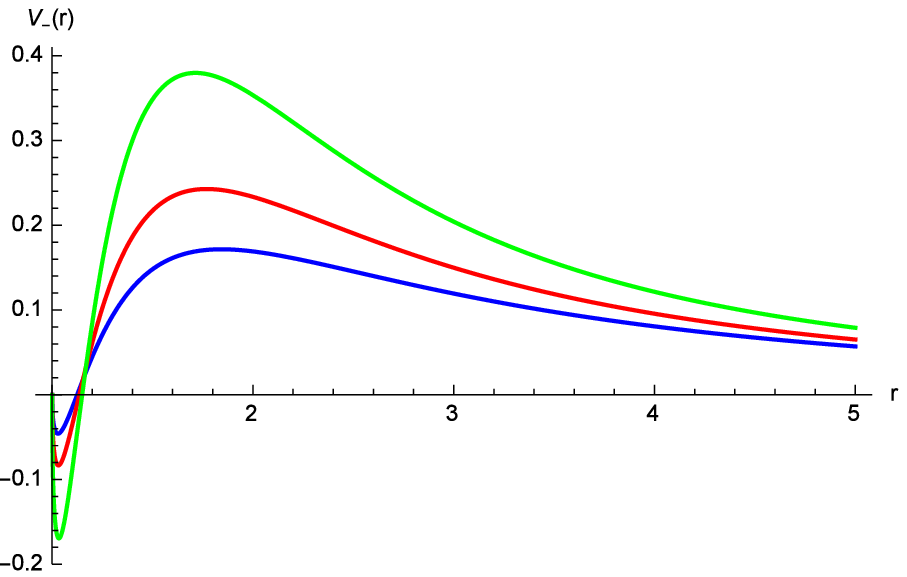} \label{fig:EW1_a} }
\hspace{4ex}
\subfigure[]{
\includegraphics[width=0.45\linewidth]{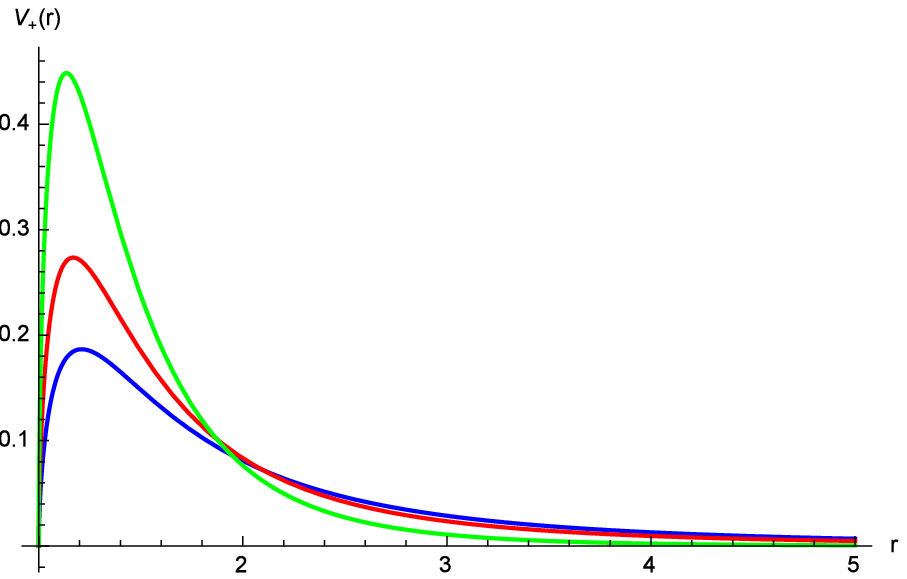} \label{fig:EW2_b} }
\caption{The effective potential $V(r)$ for the EW gravity for $\ell=1$; the blue line
corresponds to $p=0.876$, the red line corresponds to $p=0.9816$ and the green line is $p=1.14$:
  \subref{fig:EdGB1_a} $V_{-}(r)$;
 \subref{fig:EdGB2_b} $V_{+}(r)$.} \label{fig:vew}
 \label{ris:one2}
\end{figure}

In the both cases for the ``plus'' (``minus'') potential of the Dirac field $k =\ell+1$ ($k =\ell$). As can be seen from figs. (\ref{ris:one1}(a), \ref{ris:one2}(a)) the potential  $V_{-}(r)$ has a negative gap near the event horizon. The same behavior is appropriate to the potential $V_{-}(r)$ in the Schwarzschild case.  However, as it was shown earlier for black holes for which both metric coefficients are equal (like for the Schwarzschild case $e^{\mu(r)}=e^{-\nu(r)}$ \cite{potentials}), the potentials of opposite chiralities can be transformed into each another with help of the Darboux transformation. This means that from both potentials we get the same quasinormal spectrum. It allows us to ignore the negative gap of the ``minus'' potential an and talk about overall stability for the Schwarzschild case. When both metric coefficients are not the same anymore, to the best of our knowledge the iso-spectrality of both chiralities was not shown. Here we can see that the following replacements:
\begin{equation}  \label{psi}
\Psi_{+}=q (W+\dfrac{d}{dr_*}) \Psi_{-}, \quad W=\sqrt{e^{\mu(r)}}, \quad q=const;
\end{equation}
provides the Darboux transformation of equations (\ref{klein-Gordon}, \ref{psi}) for transition between ``minus'' (given by $V_{-}(r)$  and ``plus'' ($V_{+}(r)$)  perturbations. As the potential for one of the chiralities is positive definite, this immediately guarantees the stability of the Dirac field for the other chirality in both considered theories.
Therefore, we can use only stable potential. Later in the work, we will use the potential $V_{+}(r)$.

\section{Quasinormal modes of massless Dirac field for Einstein-dilaton-Gauss-Bonnet and Einstein-Weyl gravities}\label{sec:massless}

For finding quasinormal modes, it is necessary to solve the spectral problem with the appropriate boundary conditions: for a functions $\Psi$ there are only incoming waves at the horizon ($r_*\rightarrow-\infty$) and only the outgoing waves at the infinity ($r_*\rightarrow+\infty$). Quite effectively this problem can be solved using the WKB-method \cite{WKB,Matyjasek:2017psv,Konoplya:2003ii,Konoplya:2019hlu}. The advantages of this method over numerical methods is the ability to obtain low-lying quasinormal modes with sufficient accuracy automatically for a broad class of effective potentials, and, thereby, not to tailor the method for each case. The method gives good accuracy when $n \leq \ell$, where $n=0,1,2,..$ is a overtone number. The general formula for the m-order of the WKB approach can be written in form:
\begin{equation}\label{wkb}
\dfrac{i(\omega^2-V_0)}{\sqrt{-2 V_0''}}-\sum\limits_{i = 2}^{m}\Lambda_i=n+\dfrac{1}{2}.
\end{equation}

Here, the $\Lambda_i$  are the correction term of the i-th order and $\Lambda_i$ depend on the value of the potentials $V(r)$ and its derivative at the maximum, ${V_0}$ is a value of $V(r)$ in $r_{max}$ and $V_0''$ is a second derivative in $r_{max}$. But the WKB series converges only asymptotically, there is no strict criterium for evaluation of an error. The higher accuracy of the WKB approach can be achieved the averaging of the Padé approximation \cite{Matyjasek:2017psv}. We will use the fourth-order of the WKB approximation and apply further Padé expansion of the order which provides the best accuracy in the Schwarzschild limit \cite{Matyjasek:2017psv,Konoplya:2019hlu}.

\subsection{An analytical approximation in the eikonal regime}\label{sec:eikonal}

In the regime of high multipole numbers $\ell$ (eikonal regime) it is sufficient to use the first order WKB formula:
\begin{equation}\label{wkbone}
\omega=\sqrt{{V_0}-i \left(n+\frac{1}{2}\right) \sqrt{-2 {V_0''}}}.
\end{equation}
When the multipole numbers $\ell$ is high the behavior of test fields of different spin obey the same law in the dominant order and the expression for $\omega$ for the Dirac field will be identical to the formulas for other spin. For Einstein-dilaton-Gauss-Bonnet case it was found for electromagnetic field \cite{Konoplya:2019hml} for small $1/\ell$:
%
\begin{subequations}
\begin{eqnarray}\label{wedgb}
\omega_{EdGB}=\frac{2}{3 \sqrt{3}r_0} \left(\left(\ell+\frac{1}{2}\right) \left(1-0.065 p\right)-i \left(n+\frac{1}{2}\right) \left(1-0.094 p\right)\right)+\mathcal{O}(p^2,\ell^{-1}),
\end{eqnarray}
\begin{eqnarray}
r_{max}= \frac{3 r_0}{2} + 0,055 r_{0} p+\mathcal{O}(p^2,\ell^{-1}),
\end{eqnarray}
\end{subequations}
where $r_{max}$ is the position of peak of the effective potential.

For the Einstein-Weyl gravity the values of $\omega$ was found in \cite{Kokkotas:2017zwt} for small $1/\ell$, where $t=1054-1203 p$ is a deviations from the Schwarzschild branch:
%
\begin{subequations}
\begin{eqnarray}\label{wew}
\omega_{EW}=\frac{2}{3 \sqrt{3}r_0} \left(\left(\ell+\frac{1}{2}\right)\left(1-0.001308 t\right)-i \left(n+\frac{1}{2}\right) \left(1-0.002743 t\right)\right)+\mathcal{O}(t^2,\ell^{-1}),
\end{eqnarray}
\begin{eqnarray}
r_{max}= \frac{3 r_0}{2} (1+0.000393 t)+\mathcal{O}(t^2,\ell^{-2}).
\end{eqnarray}
\end{subequations}

When $p = 0$ in the formula (\ref{wedgb}) and $t = 0$ in (\ref{wew}) these formulas go over into the well-known eikonal formula for the Schwarzschild black hole. A general approach to finding eikonal quasinormal modes for static asymptotically flat and spherically symmetric black holes has been recently suggested in \cite{Churilova:2019jqx}.
It is worthwhile mentioning that the real and imaginary parts of the above eikonal formulas for test fields will coincide with the oscillation frequency and the Lyapunov exponents of the null geodesics in the background of the Einstein-dilaton-Gauss-Bonnet and Einstein-Weyl black holes \cite{Cardoso:2008bp}. However, this is not expected for the gravitational or other non-test (non-minimally coupled) fields \cite{Konoplya:2017wot,Breton:2017hwe,Toshmatov:2018ell}.

\subsection{Quasinormal modes for low $\ell$}\label{sec:lsmall}

\begin{figure}[ht]
\vspace{-4ex} \centering \subfigure[]{
\includegraphics[width=0.29\linewidth]{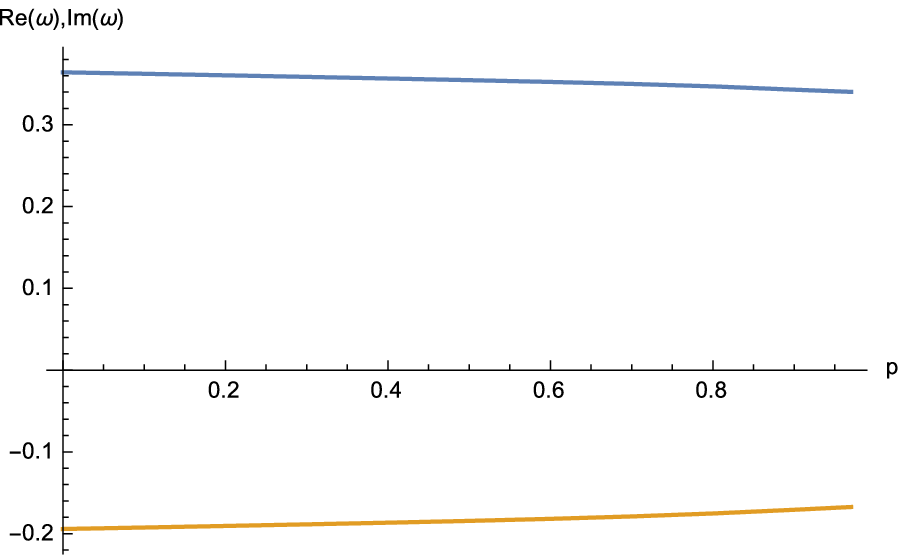} \label{fig:EdGBl1_a} }
\hspace{4ex}
\subfigure[]{
\includegraphics[width=0.29\linewidth]{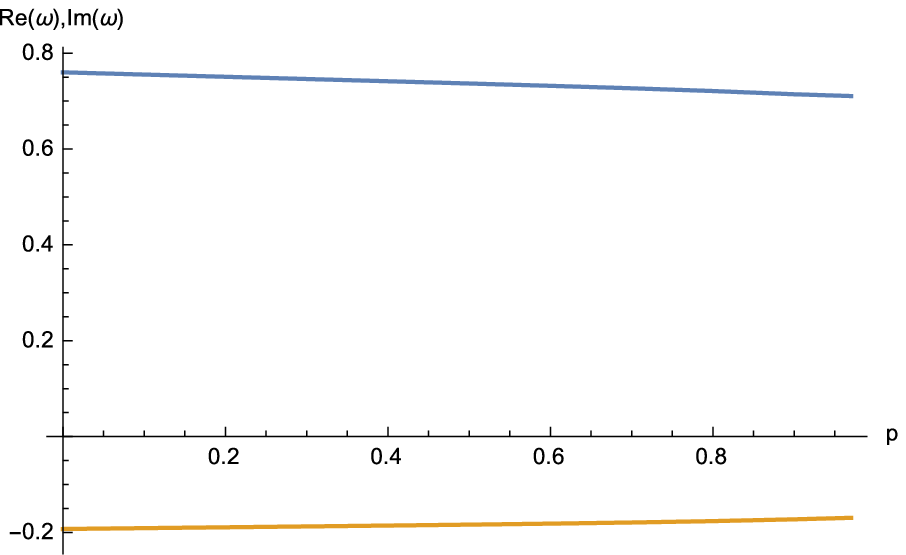} \label{fig:EdGBl2_b} }
\hspace{4ex}
\subfigure[]{ \includegraphics[width=0.29\linewidth]{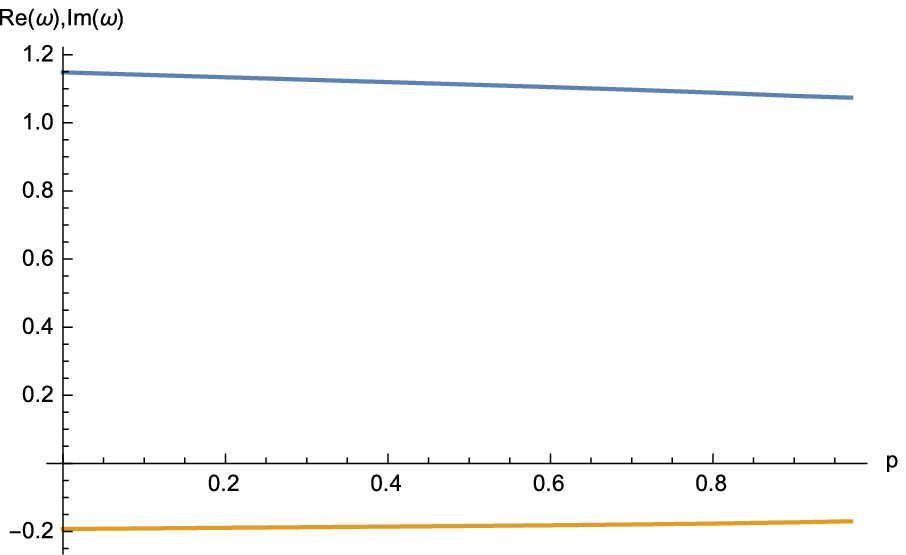} \label{fig:EdGBl3_c} }
\caption{The fundamental quasinormal mode of EdGB ($n=0$) for the Dirac field ($s=1/2$), blue line is real part of frequency, red line is imaginary part;  positive values for the real part and negative values for the imaginary part:
 \subref{fig:EdGBl1_a} $\ell = 1$;
 \subref{fig:EdGBl2_b} $\ell = 2$;
 \subref{fig:EdGBl3_c} $\ell = 3$.} \label{fig:qnmedgb}
 \label{ris:one}
\end{figure}

\begin{figure}[ht]
\vspace{-4ex} \centering \subfigure[]{
\includegraphics[width=0.29\linewidth]{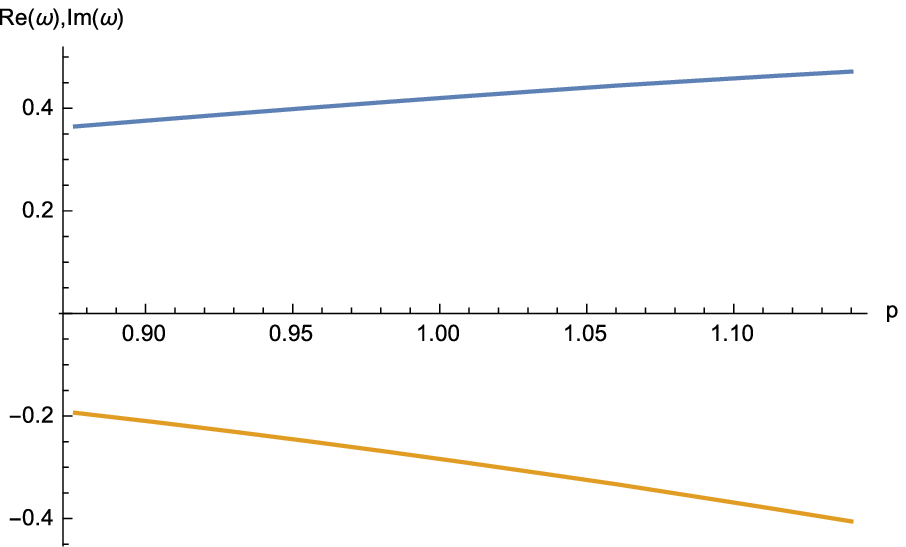} \label{fig:EWl1_a} }
\hspace{4ex}
\subfigure[]{
\includegraphics[width=0.29\linewidth]{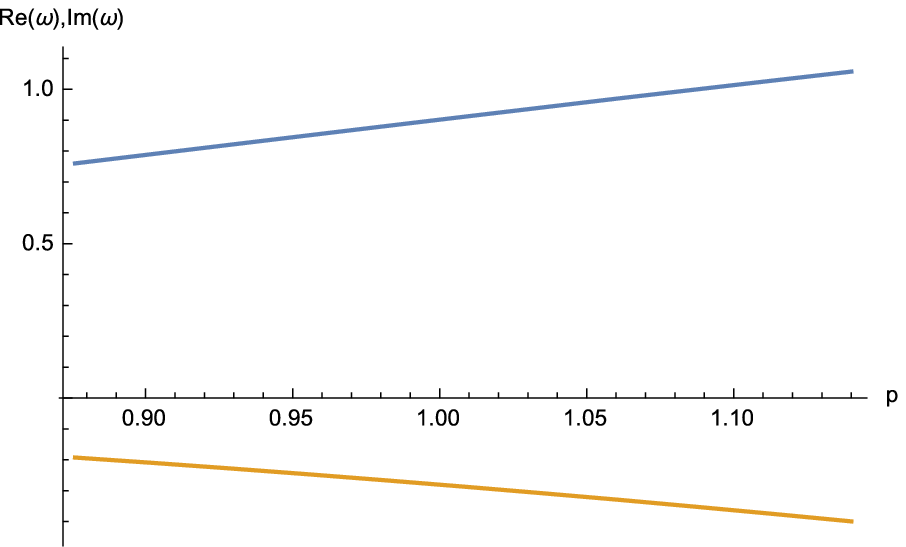} \label{fig:EWl2_b} }
\hspace{4ex}
\subfigure[]{ \includegraphics[width=0.29\linewidth]{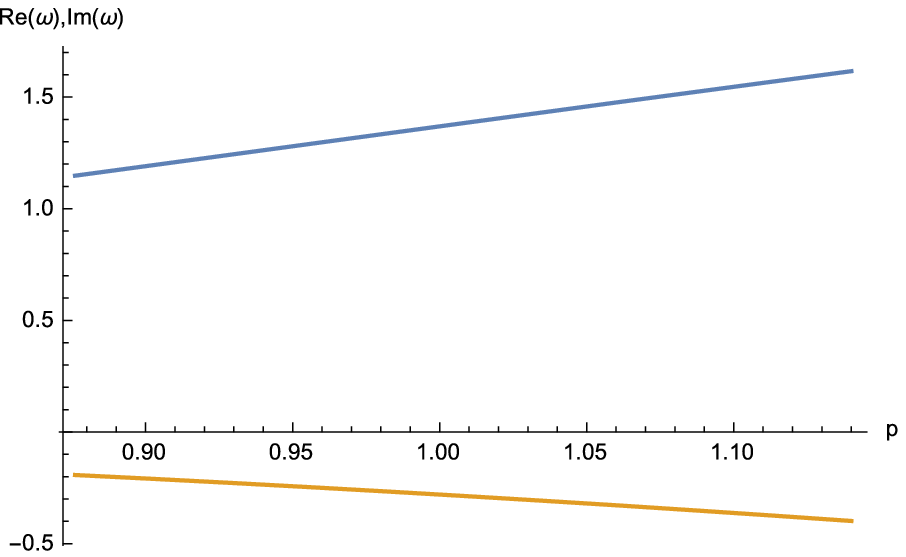} \label{fig:EWl3_c} }
\caption{The fundamental quasinormal mode of EW ($n=0$) for the Dirac field ($s=1/2$), blue line is real part of frequency, red line is imaginary part;  positive values for the real part and negative values for the imaginary part:
 \subref{fig:EWl1_a} $\ell = 1$;
 \subref{fig:EWl2_b} $\ell = 2$;
 \subref{fig:EWl3_c} $\ell = 3$.} \label{fig:qnmew}
 \label{ris:two}
\end{figure}

For obtaining accurate values of quasinormal modes at low numbers $\ell$ we will use  the fourth-order of the WKB approximation (\ref{wkb}) and apply further Padé expansion of the order. In the figs. (\ref{ris:one}, \ref{ris:two}) we construct the real (oscillation frequency) and imaginary (damping rate of oscillation) parts of the frequency $\omega$ on the values of the parameter $p$ for various multipole numbers $\ell$. As can be seen, the function $\omega(p)$ tends to be linear for all cases. This behavior is also characteristic of other test fields that were previously considered \cite{Konoplya:2019hml}, \cite{Zinhailo:2018ska}. Comparing figs. (\ref{ris:one}, \ref{ris:two}), we can see that the deviations from Schwarzschild branch by the Weyl correction are much larger than the Einstein-dilaton-Gauss-Bonnet gravity. For the Einstein-dilaton-Gauss-Bonnet case values of the modes are decreasing when increasing the dimensionless parameter $p$. On the contrary, for Weyl case we see, that the oscillation frequency and the damping rate of oscillations are increasing with increasing $p$. This also follows from the form of curves for potentials figs. (\ref{ris:one1}, \ref{ris:one2}). With an increase the dimensionless parameter $p$ for the Einstein-dilaton-Gauss-Bonnet gravity, the height of the potential barrier decreases, which is on the favor of lower bound states. For the Einstein-Weyl gravity, the maximum of potential $V(r)$ increases with increasing $p$. It means, that with increasing $p$ the corresponding frequencies are higher.
 \vspace*{0.5em plus .6em minus .5em}

Approximate calculation formulas for the complex frequency were found from the obtained data for different values $\ell$. For Einstein-dilaton-Gauss-Bonnet it is (\ref{EdGB}):
\begin{subequations}\label{EdGB}
\begin{eqnarray}
Re(\omega_{s=0.5,\ell=1})\approx0.444-0.089 p,
\nonumber\\
Im(\omega_{s=0.5,\ell=1})\approx-0.284+0.100 p;\\
Re(\omega_{s=0.5,\ell=2})\approx0.927-0.189 p,
\nonumber\\
Im(\omega_{s=0.5,\ell=2})\approx-0.269+0.086 p;\\
Re(\omega_{s=0.5,\ell=3})\approx1.399-0.285 p,
\nonumber\\
Im(\omega_{s=0.5,\ell=3})\approx:-0.266+0.083 p.
\end{eqnarray}
\end{subequations}

Accordingly, for Einstein-Weyl $\ell = 1$,  $\ell = 2$,  $\ell = 3$, we have (\ref{EW}):
\begin{subequations}\label{EW}
\begin{eqnarray}
Re(\omega_{s=0.5,\ell=1})\approx0.011+0.407 p,
\nonumber\\
Im(\omega_{s=0.5,\ell=1})\approx0.515-0.803 p;\\
Re(\omega_{s=0.5,\ell=2})\approx-0.227+1.128 p,
\nonumber\\
Im(\omega_{s=0.5,\ell=2})\approx0.499-0.783 p;\\
Re(\omega_{s=0.5,\ell=3})\approx-0.407+1.776 p,
\nonumber\\
Im(\omega_{s=0.5,\ell=3})\approx0.496-0.780 p.
\end{eqnarray}
\end{subequations}

Approximate dependencies for parts of the complex frequency in $p$ were obtained in formulas (\ref{EdGB}) and (\ref{EW}) for low $\ell$. For all options, we have a reasonable linear approximation, which is clearly visible in figs. (\ref{ris:one}) and (\ref{ris:two}).

\subsection{The checking of Hod's conjecture}\label{sec:hods}

In work \cite{Hod:2006jw}, Hod put forward the statement for damping rate of the fundamental oscillation. In other words in the spectrum of quasinormal modes there always must exist a frequency whose absolute value of the imaginary part is smaller than $\pi$ times Hawking temperature of the black hole.  According to this statement, for asymptotically flat black holes as well as for nonasymptotically flat ones, the following inequality holds:
\begin{equation}\label{hod}
|Im(\omega)|\leq\pi T_H,
\end{equation}

where $T_H$ is the Hawking temperature. The Hawking temperature $T_H$ for Einstein-dilaton-Gauss-Bonnet and Einstein-Weyl gravities can be written in the next form:
\begin{equation}\label{hod}
T_{H}=\frac{1}{4 \pi} \sqrt{\frac{de^{\mu(r)}}{dr} \frac{de^{-\nu(r)}}{dr}}\bigg|_{r=r_{0}}.
\end{equation}

\begin{figure}[ht]
\vspace{-4ex} \centering \subfigure[]{
\includegraphics[width=0.45\linewidth]{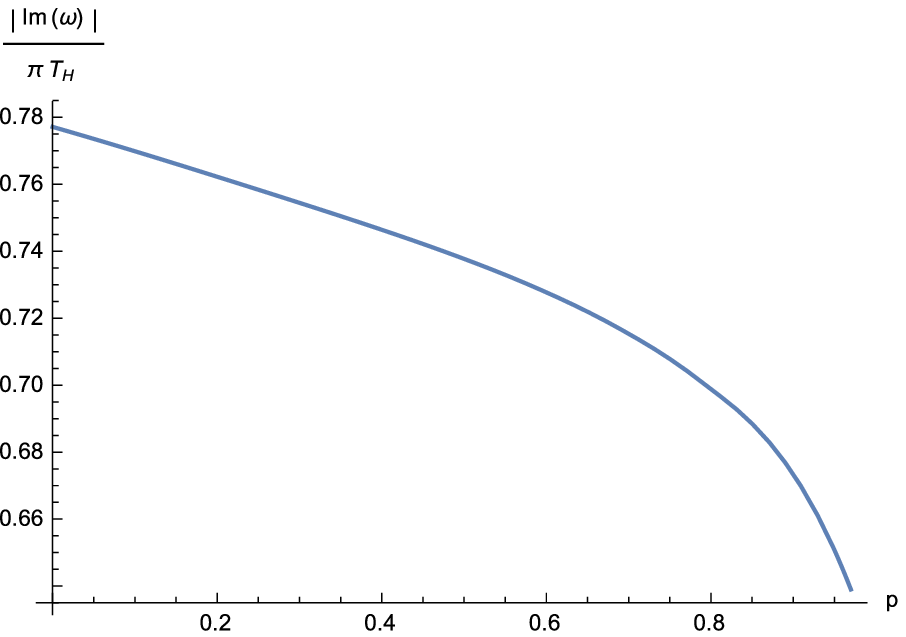} \label{fig:EdGBT1_a} }
\hspace{4ex}
\subfigure[]{
\includegraphics[width=0.45\linewidth]{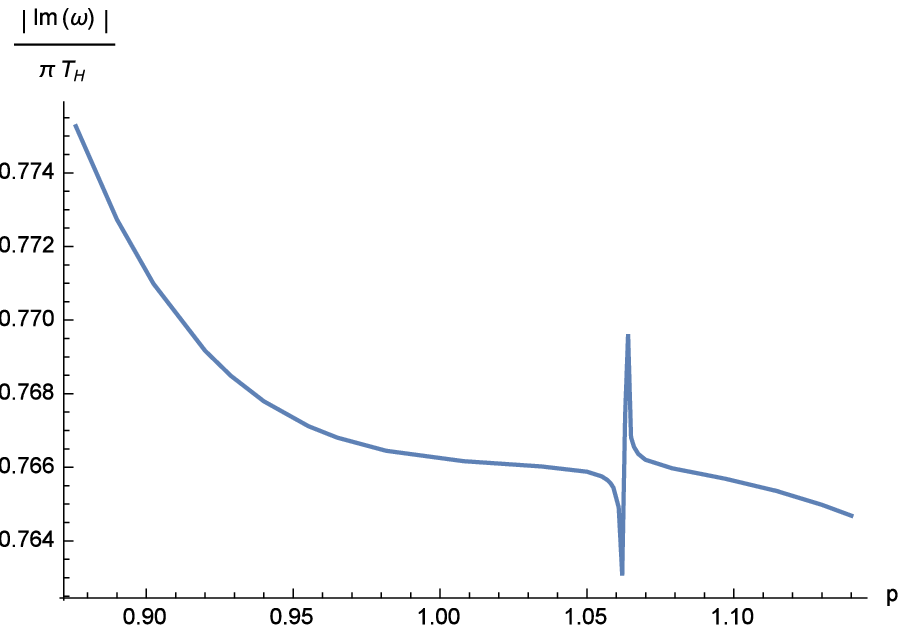} \label{fig:EWT2_b} }
\caption{The dependencies of Hod's conjecture on $p$:
 \subref{fig:EdGBT1_a} $EdGB$;
 \subref{fig:EWT2_b} $EW$.} \label{fig:th}
 \label{ris:th}
\end{figure}

From the fig. (\ref{ris:th}) it can be seen that for the whole interval of values of parameter $p$ for Einstein-dilaton-Gauss-Bonnet and Einstein-Weyl metrics $\frac{|Im(\omega)|}{\pi T_H}\leq1$. It means, that for the Hod's conjecture also holds for the cases considered.

\section{Conclusions}\label{sec:conclusions}

In this work we considered test massless Dirac field in Einstein-dilaton-Gauss-Bonnet and Einstein-Weyl gravities. It was shown that although the potential $V_{-}(r)$ of the Dirac field has a negative gap near the event horizon, we have proved that the Dirac field is stable in both considered theories. This is possible because of the stability of the second potential $V_{+}(r)$ and the iso-spectrality of both potentials. Quasinormal modes were obtained for both metrics for different values of the angular parameter $\ell$. The dependence of the complex frequency on the new parameter $p$ was constructed. The Einstein-Weyl gravity allows for much stronger deviations form the Schwarzschild geometry. Therefore, quasinormal modes of Einstein-Weyl black hole are more different from the Schwazrschild case than those of Einstein-dilaton-Gauss-Bonnet black hole, achieving the effect of tens of percents. In the last part of this work we shown, that the Hod's conjecture holds for Einstein-dilaton-Gauss-Bonnet and Einstein-Weyl gravities for the Dirac field.

In the future, it would be interesting to investigate the Dirac field including the massive term and check the possibility of the existence of the arbitrarily long-lived quasinormal modes, called quasiresonances \cite{Ohashi:2004wr}, for this case. In \cite{Konoplya:2017tvu} it has recently been shown that the quasiresonances exist for the massive Dirac field in the Einstein theory, but no such study was performed in the higher curvature corrected theories. Our approach could also be extended to the case of Einstein-Gauss-Bonnet black holes with other types of coupling of the scalar field  \cite{Konoplya:2019fpy} as well as to scalarized black holes for whose metrics analytical approximations are known \cite{Konoplya:2019goy}.

\acknowledgments{
The author acknowledges the  support of the grant 19-03950S of Czech Science Foundation ($GA\check{C}R$) and acknowledges the SU grant SGS/12/2019}. The author would like to acknowledge Roman Konoplya for useful discussions.

\newpage
\appendix
\section{Analytical form of the metric functions of Einstein-dilaton-Gauss-Bonnet metric}\label{Appendix1}
\begin{widetext}
The analytical approximations for the metric functions $e^{\mu(r)}$ and $e^{\nu(r)}$ have the forms:
\begin{subequations}
\begin{eqnarray}
e^{\mu(r)} &=&
[(r - r_{0}) (11528 (-338485+167871 p + 937132 p^2-1091895 p^3+325377 p^4) r^4+8 (263522875
\nonumber \\&&
+497564855 p-2160940683 p^2+1833700801 p^3-382791763 p^4-54635232 p^5+3579147 p^6) r^3 r_{0}
\nonumber \\&&
-124488 (-1+p)^2 p (-1310+1551 p-514 p^2+33 p^3) r^2 r_{0}^2+p (283646440-1112933120 p
\nonumber \\&&
+1868830098 p^2-1478746401 p^3+470844780 p^4-32741280 p^5) r r_{0}^3+1441 p (-234080+345600 p
\nonumber \\&&
-85004 p^2-36868 p^3+11115 p^4) r_{0}^4)]/[11528 (-1+p) (-5+3 p) r^4 ((-67697-74741 p
\nonumber \\&&
+108459 p^2) r+(36575+121424 p-124020 p^2) r_{0})],
\end{eqnarray}
\begin{eqnarray}
e^{\nu(r)} &=&
[2882 (-1+ p) (-5+3 p) r^2 ((-67697-74741 p+108459 p^2) r+(36575+121424 p
\nonumber \\&&
-124020 p^2) r_{0}) (18 (-297882+533046 p-262075 p^2+24795 p^3) r^2+18 (223782-348366 p+110455 p^2
\nonumber \\&&
+16245 p^3) r r_{0}-95 p (-3640+8312 p-6075 p^2+1404 p^3) r_{0}^2)^2]/[81 (13-9 p)^2 (r-r_{0}) ((22914
\nonumber \\&&
-25140 p+2755 p^2) r+(-17214+14880 p+1805 p^2) r_{0})^2 (11528 (-338485+167871 p+937132 p^2
\nonumber \\&&
-1091895 p^3+325377 p^4) r^4+8 (263522875+497564855 p-2160940683 p^2+1833700801 p^3
\nonumber \\&&
-382791763 p^4-54635232 p^5+3579147 p^6) r^3 r_{0}-124488 (-1+p)^2 p (-1310+1551 p-514 p^2
\nonumber \\&&
+33 p^3) r^2 r_{0}^2+p (283646440-1112933120 p+1868830098 p^2-1478746401 p^3+470844780 p^4
\nonumber \\&&
-32741280 p^5) r r_{0}^3+1441 p (-234080+345600 p-85004 p^2-36868 p^3+11115 p^4) r_{0}^4)].
\end{eqnarray}
\end{subequations}
\end{widetext}

\section{Analytical form of the metric functions of Einstein-Weyl metric}\label{Appendix2}
\begin{widetext}
The metric coefficients are determined as follows:
\begin{equation}
e^{\mu(r)} = \left(1-\frac{r_0}{r}\right)A(r), \qquad e^{\nu(r)} = \frac{B(r)^2}{\left(1-\frac{r_0}{r}\right)A(r)},
\end{equation}
where
\begin{subequations}
\begin{eqnarray}
A(r)&=&\Biggr[152124199161 \left(873828 p^4-199143783 p^3+806771764 p^2-1202612078 p+604749333\right) r^4
\\\nonumber&&+78279\left(1336094371764p^6-300842119184823 p^5+393815823540843 p^4+2680050514097926 p^3\right.
\\\nonumber&&\left.-9501392159249689 p^2+10978748485369369 p-4249747766121792\right)r^3 r_0
\\\nonumber&&-70372821 \left(1486200636 p^6+180905642811 p^5+417682197141 p^4-1208134566031 p^3\right.
\\\nonumber&&\left.-324990706209 p^2+3382539200269p-2557857695019\right) r^2 r_0^2-\left(104588131327314156 p^6\right.
\\\nonumber&&-23549620247668759617 p^5-435688050031083222417p^4+2389090517292988952355 p^3
\\\nonumber&&\left.-3731827099716921879958 p^2+2186684376605688462974 p-389142952738481370396\right) r r_0^3
\\\nonumber&&+31\left(3373810687977876 p^6+410672271594465801 p^5-14105000476530678231 p^4+51431640078486304191 p^3\right.
\\\nonumber
&&\left.-71532183052581307042p^2+43250367615320791700 p-9476049523901501640\right) r_0^4\Biggr]
\\\nonumber&&/\Biggr[152124199161 r^2\Biggr(\left(873828 p^4-199143783 p^3+806771764 p^2-1202612078 p+604749333\right) r^2
\\\nonumber&&-2 \left(873828 p^4-47583171 p^3+386036980 p^2-678598463 p+341153481\right) r r_0
\\\nonumber&&+899\left(972 p^4+115659 p^3-38596 p^2-1127284 p+1101579\right) r_0^2\Biggr)\Biggr]\,,
\end{eqnarray}
\begin{eqnarray}
B(r)&=&\Biggr[464405 \left(3251230164 p^3-14548777134 p^2+20865434326 p+23094914865\right) r^3
\\\nonumber&&-464405 \left(6502460328 p^3-52856543928p^2+100077612184 p-32132674695\right) r^2 r_0
\\\nonumber&&-\left(1244571650887908 p^3+17950319416564777 p^2-53210739821255918p+5097428297648940\right) r r_0^2
\\\nonumber
&&+635371 \left(4335198168 p^3-42352710803 p^2+90235778452 p-49464019740\right)r_0^3\Biggr]
\\\nonumber&&/\Biggr[464405 r \Biggr(\left(3251230164 p^3-14548777134 p^2+20865434326 p+23094914865\right) r^2
\\\nonumber&&-\left(6502460328p^3-52856543928 p^2+100077612184 p-32132674695\right) r r_0
\\\nonumber&&+6\left(541871694 p^3-6384627799 p^2+13202029643p+2626009760\right) r_0^2\Biggr)\Biggr]\,.
\end{eqnarray}
\end{subequations}
\end{widetext}

\end{document}